\documentclass{acm_proc_article-sp}
\usepackage{epsfig,url}

\begin{document}

\title{Enhancing Content-And-Structure Information Retrieval using a Native XML Database}

\numberofauthors{3}

\author{
%
\alignauthor Jovan Pehcevski\\
       \affaddr{RMIT University}\\
       \affaddr{Melbourne, Australia}\\
       \email{jovanp@cs.rmit.edu.au}
\alignauthor James Thom\\
       \affaddr{RMIT University}\\
       \affaddr{Melbourne, Australia}\\
       \email{jat@cs.rmit.edu.au}
\alignauthor Anne-Marie Vercoustre\\
       \affaddr{INRIA}\\
       \affaddr{Rocquencourt, France}\\
       \email{anne-marie.vercoustre@inria.fr}
}
\date{04 June 2004}
\maketitle
\begin{abstract}
Three approaches to content-and-structure XML retrieval are analysed in this paper:
first by using Zettair, a full-text information retrieval system;
second by using eXist, a native XML database, and third
by using a hybrid XML retrieval system that uses eXist to produce the final answers from
likely relevant articles retrieved by Zettair.
INEX 2003 content-and-structure topics can be classified in two categories:
the first retrieving full articles as final answers,
and the second retrieving more specific elements within articles as final answers.
We show that for both topic categories our initial hybrid system
improves the retrieval effectiveness of a native XML database.
For ranking the final answer elements,
we propose and evaluate a novel retrieval model
that utilises the structural relationships between the answer elements of a native XML database and
retrieves \emph{Coherent Retrieval Elements}.
The final results of our experiments show that
when the XML retrieval task focusses on highly relevant elements
our hybrid XML retrieval system with the Coherent Retrieval Elements module
is 1.8 times more effective than Zettair and 3 times more
effective than eXist, and yields an effective content-and-structure XML retrieval.
\end{abstract}


\section{Introduction}

This paper explores an effective hybrid XML retrieval approach that combines full-text information retrieval features
with XML-specific retrieval features found in a native XML database.
We focus on
improving XML retrieval for content-and-structure (CAS) retrieval topics, which represent
topics that enforce constraints on the existing document structure
and explicitly specify the type of the unit of retrieval (such as section or paragraph).
A retrieval challenge for a CAS topic is providing \emph{relevant} answers to a user information need.
In our previous work~\cite{KluwerINEX03} we investigated the impact of different XML retrieval approaches
on content-only (CO) retrieval topics, and also proposed a hybrid system as an effective retrieval solution.
Both CAS and CO topics are part of INEX\footnote{http://www.is.informatik.uni-duisburg.de/projects/inex/index.html.en},
the INitiative for the Evaluation of XML Retrieval.

The INEX 2003 CAS retrieval topics can be classified in
two categories: the first category of topics
where full articles rather than more specific elements are required to be retrieved as final answers, and
the second category of topics where more specific elements rather than full articles are required to be retrieved as final answers.

For topics in the first category, we investigate whether a full-text information retrieval system
is capable of retrieving full article elements as highly relevant answers.
We use Zettair\footnote{http://www.seg.rmit.edu.au/lucy/} (formerly known as Lucy)
as our choice for a full-text information retrieval system.
Zettair is a compact and fast full-text search engine designed and written by the Search
Engine Group at RMIT University.
Although Zettair implements an efficient inverted index structure~\cite{MG99},
the unit of retrieval is a full article, and currently it is neither capable of
indexing and retrieving more specific elements within articles nor capable of
specifying constraints on elements within articles.

For topics in the second category, we investigate whether an XML-specific retrieval system
is capable of retrieving more specific elements as highly relevant answers.
We use eXist\footnote{http://exist-db.org/}, an open source native XML database,
as our choice for an XML-specific retrieval system.
eXist implements many XML retrieval features found in most native XML databases,
such as full and partial keyword text searches and proximity functions.
Two of eXist's advanced features are efficient
index-based query processing and XPath extensions for full-text search~\cite{eXist}.
However, most native XML databases
follow Boolean retrieval approaches
and are not capable of
ranking the final answer elements
according to their estimated likelihood of relevance to the information need in a CAS retrieval topic.

Our initial experiments using a native XML database approach show a poor retrieval performance
for CAS retrieval topics. We also observe a similar retrieval behaviour for CO retrieval topics~\cite{LuXistINEX03, KluwerINEX03}.
In an effort to enhance its XML retrieval effectiveness, we implement
a retrieval system that follows a \emph{hybrid} XML retrieval approach.
The native XML database in our hybrid system effectively utilises the
information about articles likely to be considered \emph{relevant} to an XML retrieval topic.
In order to address the issue of ranking the final answer elements,
we develop and evaluate a retrieval module that
for a CAS topic utilises the structural relationships found in the answer list
of a native XML database and retrieves a ranked list of \emph{Coherent Retrieval Elements (CREs)}.
Section 3.4 provides the definition of CREs and highlights their importance in the XML retrieval process.
Our module can equally be applied to both cases when the logical query constraints
in a CAS topic are treated as either strict or vague, since it is capable of identifying
highly relevant answer elements at different levels of retrieval granularity.

The hybrid system and the CRE retrieval module we use in this paper extend the system and the module we previously
proposed and evaluated for the INEX 2003 CO retrieval topics~\cite{KluwerINEX03}.

\section{Analysis of INEX 2003 Topics}

INEX provides
a means, in the form of a test collection and corresponding scoring methods,
to evaluate the effectiveness of different XML retrieval systems.
INEX uses an XML document collection that comprises 12107 IEEE Computer Society
articles published in the period 1997-2002 with approximately 500MB of data.
Each year (starting in 2002) a new set of XML retrieval topics are introduced in INEX
which are then usually assessed by participating groups that originally proposed the topics.

The XML retrieval task performed by the groups participating in INEX is
defined as ad-hoc retrieval of XML documents. In information retrieval
literature this type of retrieval involves searching a static set of documents
using a new set of topics, which represents an activity commonly used in digital library systems.

Within the ad-hoc retrieval task, INEX defines two additional retrieval tasks:
a \emph{content-only (CO)} task involving CO topics,
and a \emph{content-and-structure (CAS)} task involving CAS topics.
A CAS topic enforces restrictions with respect to the
underlying document structure by explicitly specifying the type of the unit of
retrieval, whereas a CO topic has no such restriction
on the elements retrieved.
In INEX 2003, the CAS retrieval task furthermore comprises a \emph{SCAS} sub-task and a \emph{VCAS} sub-task.
A SCAS sub-task considers the structural constraints in a query to be
\emph{strictly} matched, while a \emph{VCAS} sub-task allows the structural constraints in
a query to be treated as \emph{vague} conditions.

In this paper we focus on improving XML retrieval for CAS topics, in particular using the SCAS retrieval sub-task.

\subsection{INEX CAS Topic Example}

\begin{table}[tb]
\begin{center}
\begin{tabular}{l}
\hline
$<?$xml\ version="1.0"\ encoding="ISO-8859-1"$?>$ \\
$<!$DOCTYPE\ inex\_topic\ SYSTEM\ "topic.dtd"$>$ \\
\ \\
$<$inex\_topic\ topic\_id="86"\ query\_type="CAS"\ ct\_no="107"$>$ \\
\ \\
$<$title$>$ \\
\ \ \ \ \ \ \ \ //sec[about(., 'mobile electronic payment system')] \\
$<$/title$>$ \\
\ \\
$<$description$>$ \\
\ \ \ \ \ \ \ \ Find sections that describe technologies for \\
\ \ \ \ \ \ \ \ wireless mobile electronic payment systems \\
\ \ \ \ \ \ \ \ at consumer level. \\
$<$/description$>$ \\
\ \\
$<$narrative$>$ \\
\ \ \ \ \ \ \ \ To be relevant, a section must describe security-\\
\ \ \ \ \ \ \ \ related technologies that exist in electronic \\
\ \ \ \ \ \ \ \ payment systems that can be implemented in \\
\ \ \ \ \ \ \ \ hardware devices. A section should be considered \\
\ \ \ \ \ \ \ \ irrelevant if it describes systems that are\\
\ \ \ \ \ \ \ \ designed to be used in a PC or laptop. \\
$<$/narrative$>$ \\
\ \\
$<$keywords$>$ \\
\ \ \ \ \ \ \ \ mobile, electronic payment system, \\
\ \ \ \ \ \ \ \ electronic wallets, e-payment, e-cash, \\
\ \ \ \ \ \ \ \ wireless, m-commerce, security \\
$<$/keywords$>$ \\
\ \\
$<$/inex\_topic$>$ \\
\hline
\end{tabular}
\end{center}
\caption{INEX 2003 CAS Topic 86}
\label{fig-INEX86}
\end{table}

Table~\ref{fig-INEX86} shows the CAS topic 86 proposed by our participating group in INEX 2003.
This topic searches for
elements within articles (sections) focusing on electronic payment technologies
implemented in mobile computing devices, such as mobile phones or handheld
devices. A section element is considered relevant if it describes
technologies that can be used to securely process electronic payments in
mobile computing devices.

Thus for a section element to be
considered marginally, fairly or highly relevant, it is very likely that it will \emph{at least}
contain a  combination of some important words or phrases, such as
\emph{mobile}, \emph{security}, \emph{electronic payment system} or
\emph{e-payment}.
Furthermore, for the INEX XML document collection the \verb+sec+, \verb+ss1+ and \verb+ss2+
elements are considered equivalent and interchangeable for a CAS topic.
In that sense,
an XML retrieval
system should follow an effective \emph{extraction strategy} capable of
producing \emph{coherent} answers with the appropriate level of retrieval granularity
(such as retrieving \verb+sec+ rather than \verb+ss2+ elements).

\subsection{INEX CAS Topic Categories}

INEX 2003 introduces 30 CAS topics in total, with topic numbers between 61 and 90.
Out of the CAS topic titles, we distinguish two categories of topics.

\begin{itemize}

\item The first category of topics seek to retrieve full articles rather than more
specific elements within articles as final answers.
There are 12 such topics, their numbers being 61, 62, 63, 65, 70, 73, 75, 79, 81, 82, 87, 88.
We refer to such topics as \emph{Article} topics.

\item The second category of topics seek to retrieve more specific
elements within articles rather than full articles as final answers.
There are 18 topics that belong to this category.
We refer to such topics as \emph{Specific} topics.

\end{itemize}

\section{XML Retrieval Approaches}

Most full-text information retrieval systems ignore the information about the document structure
and consider whole documents as units of retrieval.
Such retrieval systems take queries that often represent a bag of words, where phrases or logical query operators could also be included.
The final list of answer elements usually comprises ranked list of whole documents sorted in a descending order
according to their estimated likelihood of relevance to the information need in the query.
Accordingly, it is expected that for CAS retrieval topics in the first category
a full-text information retrieval system would be able to successfully retrieve highly relevant articles.

Most native XML databases support XML-specific retrieval technologies, such as found in XPath and XQuery.
The information about the structure of the XML documents is usually incorporated in the document index, allowing
users to query both by document content and by document structure.
This allows an easy identification of elements that belong to the XML documents,
either by the path they appear in the document or by certain keywords they contain.
Accordingly, it is expected that
a native XML database would be suitable for CAS retrieval topics that belong in the second category.

In an effort to support a content-and-structure XML retrieval that combines both CAS topic categories,
we develop a \emph{hybrid} XML retrieval system that
uses a native XML database to produce final answers from those documents that are estimated as likely to be relevant
by a full-text information retrieval system.

The following sections describe the XML retrieval approaches implemented in the respective systems,
together with some open issues that arise when a particular retrieval approach is applied.

\subsection{Full-Text Information Retrieval Approach}

The efficient inverted index structure is first used with Zettair
to index the INEX XML document collection.
The term postings file is stored in a compressed form on disk, so
the size of the Zettair index takes roughly
26\% of the total collection size.
The time taken to index the entire INEX collection on a system with a Pentium4 2.66GHz processor and a 512MB RAM memory
running Mandrake Linux 9.1 is around 70 seconds.

A topic translation module is used to
automatically translate an INEX CAS topic into a Zettair query.
For INEX CAS topics, terms that
appear in the \verb+<Title>+ part of the topic are used to formulate the query.
Up to 100 \verb+<article>+ elements are then returned
in the descending order according to their estimated likelihood of relevance to the CAS topic.
One retrieval issue when using Zettair, which is in particular related to the XML retrieval process,
is that it is not currently capable of indexing and retrieving more specific elements within articles.

When the information retrieval task involves retrieval of whole documents with varying lengths,
the pivoted cosine document length normalisation scheme is shown to be an effective retrieval scheme~\cite{Pivot}.
For the INEX XML document collection, we calculated the optimal slope parameter in the pivoted cosine ranking formula
by using a different set of retrieval topics (those from the previous year, INEX 2002).
When using terms from \verb+<Title>+ part of INEX topics while formulating Zettair queries,
we found that a slope parameter with a value of 0.25 yields highest system effectiveness
(although when longer queries are used, such as
queries that contain terms from the \verb+<Keywords>+ part of INEX topics,
a different value of 0.55 would be better~\cite{KluwerINEX03}).
Consequently, for INEX 2003 CAS topics
we use the value of 0.25 for the slope parameter in the pivoted cosine ranking formula in Zettair.

\subsection{Native XML Database Approach}

With eXist, the INEX XML document collection is first indexed by using its efficient indexing scheme.
This index stores the information about the parsed elements within articles together
with the information about the attributes and all word occurrences; its size is
roughly twice as big as the total collection size.
The time taken to index the entire INEX collection on a system with a Pentium 4 2.6GHz processor and a 512MB RAM memory
running Mandrake Linux 9.1 is around 2050 seconds.

\begin{figure*}[tb]
\begin{center}
\epsfxsize=12cm
\centering\epsfbox{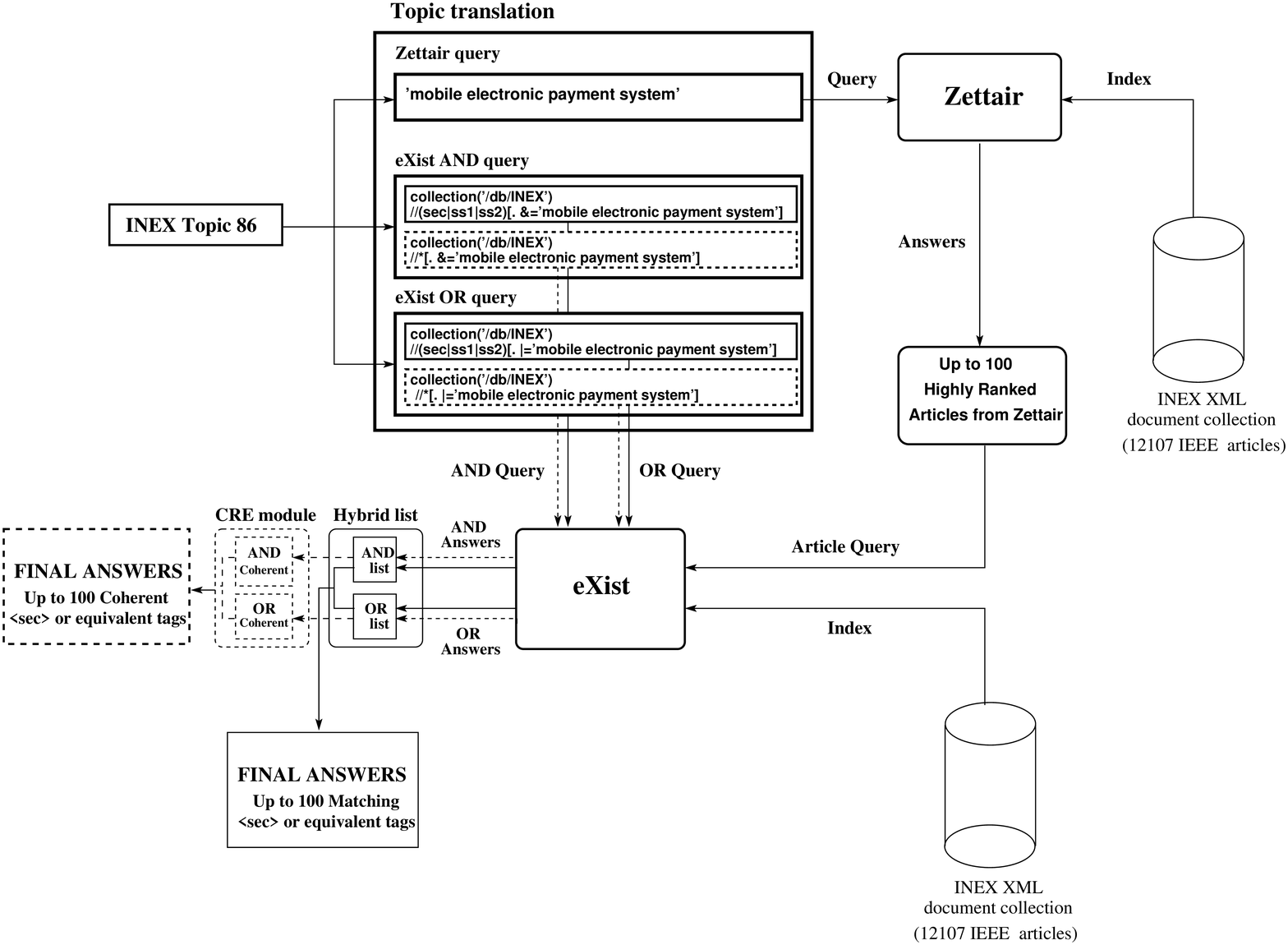}
\caption{A hybrid XML retrieval approach to INEX CAS topics.}
\label{fig-hybrid}
\end{center}
\end{figure*}

A topic translation module is used to
automatically translate an INEX CAS topic into two eXist queries: AND and OR.
For INEX CAS topics, the terms and structural constraints that
appear in the \verb+<Title>+ part of the CAS topic are used to formulate eXist queries.
The \verb+&=+ and \verb+|=+ query operators are used with eXist
while formulating the above queries, respectively.
The AND and OR eXist queries are depicted in solid boxes in Figure~\ref{fig-hybrid}
where the elements to be retrieved are specified explicitly.

For an INEX CAS topic, our choice for the final list of answer elements comprises matching elements from the AND answer list
followed by the matching elements from the OR answer list that do not belong to the AND answer list.
If an AND answer list is empty, the final answer list is the same as the OR answer list.
In both cases it contains (up to) 100 matching articles or elements within articles.
The equivalent matching elements are also considered during the retrieval process.

We observed two retrieval issues while using eXist,
which are in particular related to the XML retrieval process.

\begin{enumerate}

\item For an INEX CAS topic that retrieves full articles rather than more specific elements within articles,
the list of answer elements comprises full articles that
satisfy the logical query constraints. These articles are sorted by their internal identifiers
that correspond to the order in which each article is stored in the database.
However, there is no information about
the \emph{estimated likelihood of relevance} of a particular matching article to
the information need expressed in the CAS topic.

\item For an INEX CAS topic that retrieves more specific elements within articles rather than full articles,
the list of answer elements comprises most specific elements that satisfy both the content and the granularity constraints in the query.
eXist orders the matching elements in the answer list by the article where they belong,
according to the XQuery specification\footnote{http://www.w3.org/TR/xquery/\#N10895}.
However, there is no information whether a particular matching element in the above list is likely
to be more relevant than other matching elements that belong to the same article.
Accordingly, \emph{ranking} of matching elements within articles is also not supported.

\end{enumerate}

The following sections describe our approaches that address both of these issues.

\subsection{Hybrid XML Retrieval Approach}

Our hybrid system incorporates the best retrieval features from Zettair and eXist.
Figure~\ref{fig-hybrid} shows the hybrid XML retrieval approach
as implemented in the hybrid system.
We use the CAS topic 86 throughout the example.
Zettair is first used to obtain (up to) 100 articles likely to be considered relevant to the information need
expressed in the CAS topic as into a Zettair query.
For each article in the answer list produced by Zettair,
both AND and OR queries are then applied by eXist,
which produce matching elements in two corresponding answer lists.
The answer list for an INEX CAS topic and \emph{a particular article} thus
comprises the article's matching elements from the AND answer list
followed by the article's matching elements from the OR answer list that do not belong to the AND answer list.

The final answer list for an INEX CAS topic comprises (up to) 100 matching elements
and equivalent element tags that belong to highly ranked articles as estimated by Zettair.
The final answer list is shown as \verb+Hybrid list+ in Figure~\ref{fig-hybrid}.

Figure~\ref{fig-hybrid} also shows queries and other parts of our hybrid system depicted in dashed boxes,
where we also explore whether using CO-type queries could improve the
CAS retrieval task.
This can equally be applied to the hybrid approach as well as to the native XML database approach,
since they both use eXist to produce the final list of matching elements.
The next section explores this retrieval process in detail.

The hybrid XML retrieval approach addresses the first retrieval issue
observed in a native XML database approach.
However, because of its modular nature we observe a loss in efficiency.
For a particular CAS topic, up to 100 articles firstly need to be retrieved by Zettair.
This article list is then queried by eXist, one article at a time.
In order to retrieve (up to) 100 matching elements, eXist may need to query each article in the list before it reaches this number.
Obviously, having an equally effective system that produces its final list of answer elements much faster
would be more efficient solution.
The second retrieval issue observed in a native XML database approach still remains open,
since for a particular article our hybrid XML retrieval system also uses eXist to produce
its final list of answer elements.

The following section describes one possible approach that addresses this issue.

\subsection{Rank the Native XML Database Output}

This section describes our novel retrieval module that
utilises the structural relationships between elements in the eXist's answer list
and identifies, ranks and retrieves Coherent Retrieval Elements.
Our definition of a Coherent Retrieval Element is as follows.

\begin{table}[tb]
\begin{center}
\begin{tabular}{l l}
\hline
Article & Answer element \\
\hline
ic/2000/w6074 & /article[1]/bdy[1]/sec[1]/ip1[1] \\
ic/2000/w6074 & /article[1]/bdy[1]/sec[1]/p[2] \\
ic/2000/w6074 & /article[1]/bdy[1]/sec[2]/ip1[1] \\
ic/2000/w6074 & /article[1]/bdy[1]/sec[2]/p[2] \\
ic/2000/w6074 & /article[1]/bdy[1]/sec[2]/p[5] \\
ic/2000/w6074 & /article[1]/bdy[1]/sec[2]/p[6] \\
ic/2000/w6074 & /article[1]/bdy[1]/sec[3]/ip1[1] \\
ic/2000/w6074 & /article[1]/bdy[1]/sec[3]/p[2] \\
ic/2000/w6074 & /article[1]/bdy[1]/sec[4]/p[3] \\
\hline
\end{tabular}
\end{center}
\caption{eXist OR answer list example}
\label{fig-eXistOR}
\end{table}

``For a particular article in the final answer list, a \emph{Coherent Retrieval Element (CRE)} is an element that contains \emph{at least} two matching elements, [or \emph{at least}] two other Coherent Retrieval Elements, or a combination of a matching element and a Coherent Retrieval Element. In all three cases, the containing elements of a Coherent Retrieval Element should constitute either its \emph{different children} or each different child's \emph{descendants}''~\cite{KluwerINEX03}.

\begin{figure}[tb]
\epsfxsize=8cm
\centering\epsfbox{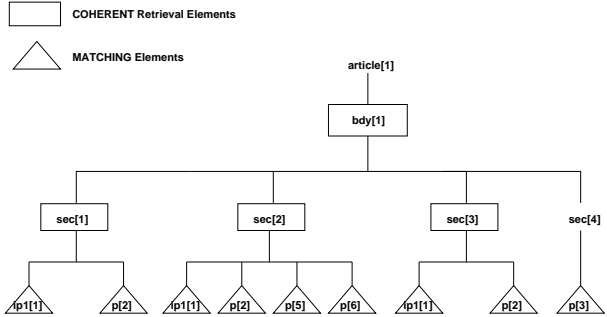}
\caption{MATCHING Elements versus COHERENT RETRIEVAL Elements: a tree-view example.}
\label{fig-coherent-tree}
\end{figure}

Consider the eXist answer list shown in Table~\ref{fig-eXistOR}.
The list is a result of using the CAS topic 86 and the OR eXist query depicted as dashed box in Figure~\ref{fig-hybrid}.
Each matching element in the list therefore contains \emph{any} combination of query keywords.
Although this example shows a retrieval case when an OR list is used with eXist,
our CRE algorithm equally applies in the case when an AND list is used.
Table~\ref{fig-eXistOR} also shows that the matching elements in the answer list are presented in article order.

Figure~\ref{fig-coherent-tree} shows a tree representation of the above eXist answer list.
The eXist matching elements are shown in triangle boxes, while the CREs are shown in square boxes.
The figure also shows elements that represent neither matching elements nor CREs.

We identify one specific case, however.
If an answer list contains only one matching element, the above CRE algorithm produces the same result: the matching element. This
is due to the lack of supporting information that would justify the choice for the ancestors of the matching element to be regarded as CREs.

So far we have managed to identify the Coherent Retrieval Elements from the eXist's answer list of matching elements to the CAS topic.
However, in order to enhance the effectiveness of our retrieval module we still need to \emph{rank}
these elements according to their \emph{estimated likelihood of relevance} to the information need expressed in the topic.
The following XML-specific heuristics are used to calculate the ranking values of the CREs (in a descending order of importance):

\begin{table}[tb]
\begin{center}
\begin{tabular}{l l c}
\hline
Article & Answer element & Rank \\
\hline
ic/2000/w6074 & /article[1]/bdy[1]/sec[2] & 1\\
ic/2000/w6074 & /article[1]/bdy[1]/sec[1] & 2\\
ic/2000/w6074 & /article[1]/bdy[1]/sec[3] & 3\\
\hline
\end{tabular}
\end{center}
\caption{Ranked list of Coherent Retrieval Elements}
\label{fig-CoherentOR}
\end{table}

\begin{enumerate}
\item The number of times a CRE appears in the absolute path of each matching element in the answer list (the more often it appears, the better);
\item The length of the absolute path of a CRE (the shorter it is, the better);
\item The ordering of the XPath sequence in the absolute path of a CRE (nearer to beginning is better); and
\item Since we are dealing with CAS retrieval task, only CREs that satisfy the granularity constraints in a CAS topic
will be considered as answers.
\end{enumerate}

In accordance to the above heuristics, if two Coherent Retrieval Elements appear the same number of times in the answer list,
the shorter one will be ranked higher.
Moreover, if they have the same length, the ordering sequence where they appear in the article will determine their final ranks.
In our example, \verb+article[1]/bdy[1]/sec[1]+ will be ranked higher than \verb+article[1]/bdy[1]/sec[3]+.

The order of importance for the XML-specific heuristics outlined above is based on the following observation.
As it is currently implemented, less specific (or more general) CREs
are likely to be ranked higher than more specific (or less general) CREs.
However, depending on the retrieval task, the retrieval module could easily be switched the other way around.
When dealing with the INEX test collection, the latter functionality proved to be less effective than the one currently implemented.

\begin{table*}[tp]
\begin{center}
\begin{tabular}{c c c c c c c}
\hline
 Quantisation & Maximum & \multicolumn{1}{c}{eXist} & \multicolumn{1}{c}{eXist-CRE} & \multicolumn{1}{c}{Hybrid} & \multicolumn{1}{c}{Hybrid-CRE} & \multicolumn{1}{c}{Zettair} \\
 \cline{3-3} \cline{4-4} \cline{5-5} \cline{6-6} \cline{7-7}
function & retrieved & matching & Coherent & matching & Coherent & full article \\
 (in inex\_eval) & elements & elements & elements & elements & elements & elements \\
 & (per article) & & & & & \\
\hline \hline
strict & 100 & 0.0682 & 0.0757 & 0.1926 & 0.2304 & 0.1264 \\
generalised & 100 & 0.0625 & 0.0588 & 0.1525 & 0.1465 & 0.0939 \\
\hline
\end{tabular}
\end{center}
\caption{Average Precision values for different XML retrieval approaches.
The values are generated by using different quantisation function in the inex\_eval evaluation metric.}
\label{LuXist-table}
\end{table*}

Table~\ref{fig-CoherentOR} shows the final ranked list of Coherent Retrieval Elements for the particular article.
(the OR list is shown in the \verb+CRE module+ in Figure~\ref{fig-hybrid}).
The \verb+bdy[1]+ element does not satisfy the last heuristic above, thus it is not included in the final list of CREs.
This means that our CRE module could easily be applied without any modifications
with the VCAS retrieval task, where the query constraints are treated as vague conditions.
Moreover, the \verb+sec[4]+ element will be included in the eXist's list of matching elements
when the strict OR query is used (the OR list is shown in the \verb+Hybrid list+ in Figure~\ref{fig-hybrid}) whereas this element
does not appear in the final list of CREs, which on the basis of the above heuristics makes it not likely to be a highly relevant element.
In that regard, we identify the Coherent Retrieval Elements as \emph{preferable units of retrieval} for the INEX CAS retrieval topics.

We show the positive impact on the XML retrieval effectiveness for the systems that implement our CRE module in the next section.

\section{Experiments and Results}

This section shows the experimental results for the above XML retrieval approaches
when different quantisation functions and different categories of CAS retrieval topics apply.
Our aim is to determine the most effective XML retrieval approach among the following:

\begin{itemize}

\item a full-text information retrieval approach, using Zettair only;
\item a native XML database approach, using eXist only;
\item a hybrid approach to XML retrieval, using our initial hybrid system;
\item a native XML database approach with the CRE module applied on the answer list; and
\item a hybrid XML retrieval approach with the CRE module applied on the answer list.

\end{itemize}

For each of the retrieval approaches above, the final answer list for an INEX CAS topic comprises
(up to) 100 articles or elements within articles.
An average precision value over 100 recall points is firstly calculated.
These values are then averaged over all CAS topics,
which produces the final average precision value for a particular retrieval run.
A retrieval run for each XML retrieval approach therefore comprises answer lists for all CAS topics.

\subsection{Comparison of XML Retrieval Approaches}

The strict quantisation function in the \verb+inex_eval+
evaluation metric~\cite{INEX02_overview}
is used to evaluate whether an XML retrieval approach is capable of retrieving highly
relevant elements.
Table~\ref{LuXist-table} shows that in this case our hybrid system that uses the CRE retrieval module is the most effective.
On the other hand, the plain eXist database is the least effective XML retrieval system.
The two previously observed retrieval issues in the native XML database approach
are very likely to influence the latter behaviour.
We also observe an 11\% relative improvement for the retrieval effectiveness
when our CRE module is applied with eXist.
Moreover, our initial hybrid system (without the CRE module)
improves the retrieval effectiveness of the plain eXist database by 2.8 times.
The latter behaviour is strongly influenced by
the presence of a full-text information retrieval system in our hybrid system.
Similar improvement for the retrieval effectiveness is exhibited
when both eXist and the hybrid system use the CRE module.

\begin{figure}[tb]
\epsfxsize=12cm
\centering\epsfbox{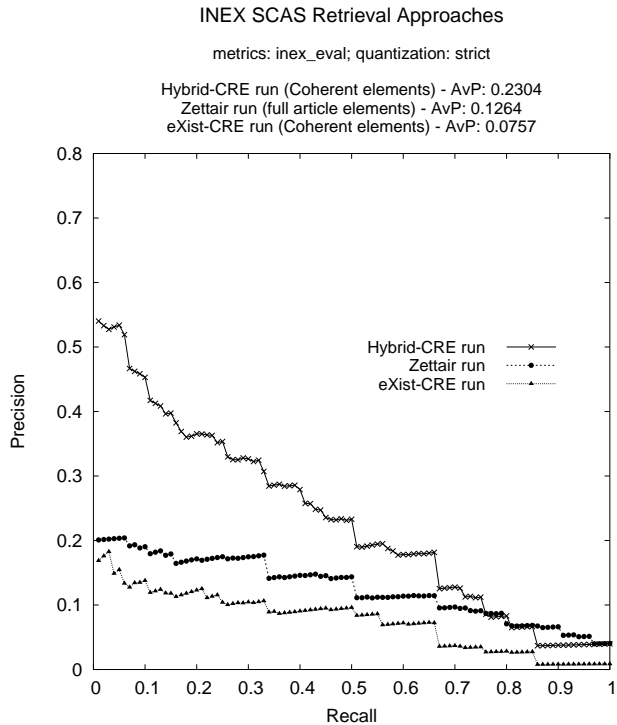}
\caption{Evaluation of the XML retrieval approaches by using strict quantisation function in inex\_eval.}
\label{fig-inex-eval}
\end{figure}

The generalised quantisation function
is used to evaluate the XML retrieval approaches
when retrieving elements with
different \emph{degrees of relevance}~\cite{INEX02_overview}.
Table~\ref{LuXist-table} shows that in this case the plain hybrid system (without the CRE module applied) performs best.
We furthermore observe that the effectiveness of retrieval systems that use the CRE module
is lower than the effectiveness of the same systems without the CRE module applied.
It is very likely that some marginally relevant elements are omitted from the list of the resulting CREs,
which, as shown before, is not the case when the XML retrieval task focusses on highly relevant elements.

Although Zettair alone can not be applied to the CAS retrieval topics in the second category,
Table~\ref{LuXist-table} shows that overall it still performs better than eXist, regardless of which quantisation function applies.
This is rather surprising, and reflects our previous expectation that
for the CAS topics in the first category Zettair is indeed capable of retrieving highly relevant articles,
whereas the first retrieval issue observed in eXist has a negative impact on its overall effectiveness.
On the other hand, both the plain hybrid system and the hybrid system with the CRE module
furthermore improve Zettair's retrieval effectiveness in either case when strict or generalised quantisation function applies.

The graph in Figure~\ref{fig-inex-eval}
outlines a detailed summary of the evaluation results for the XML retrieval approaches
when the standard \verb+inex_eval+ evaluation metric using strict quantisation function applies.
It shows runs that produce the best results for each XML retrieval approach, which
(except plain Zettair) represent the approaches that apply the CRE retrieval module.
As previously observed, the hybrid-CRE run performs best, followed by the Zettair run, and the eXist-CRE run is worst.

\subsection{Analysis based on CAS Topic Categories}

Our last experiment is based upon the INEX 2003 CAS topic categories described in Section 2.2.
The retrieval effectiveness of the XML retrieval approaches is evaluated across three CAS topic categories:
\emph{All} topics, \emph{Article} topics and \emph{Specific} topics.
The strict quantisation function in \verb+inex_eval+ evaluation metric is used
to calculate the average precision values for each run.

\begin{table}
\begin{center}
\begin{tabular}{lccc}
\hline
XML retrieval & \multicolumn{3}{c}{INEX 2003 CAS Topics} \\
\cline{2-4}
approach & $~All$ & $~Article$ & $~Specific$ \\
\hline
\hline
$~Hybrid-CRE$ & 0.2304 & 0.2636 & 0.2083 \\
$~Zettair$ & 0.1264 & 0.3144 & 0.0000 \\
$~eXist-CRE$ & 0.0757 & 0.0736 & 0.0771 \\
\hline
\end{tabular}
\end{center}
\caption{Average precision values for the XML retrieval approaches over different CAS topic categories.
The values are generated by using strict quantisation function in inex\_eval.}
\label{topic-categories}
\end{table}

Table~\ref{topic-categories} shows final results for each XML retrieval approach evaluated across the three topic categories.
For Article topics the Zettair run performs best,
outperforming the hybrid-CRE run and the eXist-CRE run.
This is very surprising, and shows that there are cases where
a highly relevant article does not necessarily represent
a matching article satisfying logical query constraints.
For Specific topics where Zettair run does not apply,
the hybrid-CRE run is roughly 2.7 times more effective than eXist-CRE run.
As shown previously, when both CAS topic categories are considered (the case of All topics), the hybrid-CRE run performs best.

\section{Related work}

Various XML retrieval approaches were used by the participating groups in INEX 2003.
These approaches were generally classified as model-oriented and system-oriented~\cite{INEX03_overview}.
Our group followed the latter approach by using the initial hybrid XML retrieval system~\cite{LuXistINEX03}.
In an earlier work regarding retrieval from semi-structured documents, Wilkinson ~\cite{RossW} shows that
simply extracting components from documents likely considered to be relevant to the information need in a query leads to poor system
effectiveness.
However, in our INEX 2003 approach we have investigated various extraction strategies with eXist
that produced effective results for CAS topics.
The hybrid system with our CRE module (which we developed since INEX 2003)
furthermore increases the retrieval effectiveness of the initial hybrid XML retrieval system.

The CSIRO group participating in INEX 2002 proposed a similar XML retrieval approach where
PADRE, the core of CSIRO's Panoptic Enterprise
Search Engine\footnote{http://www.panopticsearch.com} is used
to rank full articles and elements within articles~\cite{INEX02_AnneMarie}.
Unlike many full-text information retrieval systems,
PADRE combines full-text and metadata indexing and retrieval and is also capable of indexing and retrieving
more specific elements within articles.
A post processing module is then used to extract and re-rank the full articles
and elements within articles returned by PADRE.
However, unlike our CRE retrieval module,
the above approach ignores the structural elements within articles that contain the indexed element.
Less specific and more general elements are therefore not likely to appear in the final answer list.

For the purpose of ranking the resulting answers of XML retrieval topics,
Wolff et al~\cite{wolff99-xpres} extend the probabilistic ranking
model by incorporating the notion of ``structural roles'', which
can be determined manually from the document schema.
However, the term frequencies are measured only for the structural elements belonging
to a particular role, without taking into account the entire context
where all these elements belong in the document hierarchy.
XRank~\cite{XRank} and XSearch~\cite{XSearch} furthermore
aim at producing effective ranked results for XML queries.
XRank generally focuses on hyperlinked XML documents, while XSearch retrieves answers
comprising semantically related nodes.
However, since the structure of IEEE XML documents in the INEX document collection does not typically meet the above requirements,
neither of them (without some modifications) could be used in a straightforward fashion with the CAS retrieval task.

\section{Conclusion and Future Work}

This paper investigates the impact when three systems with
different XML retrieval approaches are used in the XML content-and-structure (CAS) retrieval task:
Zettair, a full-text information retrieval system; eXist, a native XML data\-base, and a hybrid XML retrieval system
that combines the best retrieval features from Zettair and eXist.

Two categories of CAS retrieval topics
can be identified in INEX 2003:
the first category of topics
where full article elements are retrieved,
and the second category of topics where
more specific elements within articles are retrieved.
We have shown that a full-text information retrieval system
yields effective retrieval for CAS topics in the first category.
For CAS topics in the second category we have used a native XML database
and have observed two issues particularly related to the XML retrieval process
that have a negative impact on its retrieval effectiveness.

In order to address the first issue as well as
support a CAS XML retrieval that combines both topic categories,
we have developed and evaluated a \emph{hybrid} XML retrieval system
that uses eXist to produce final answers from the likely relevant articles retrieved by Zettair.
For addressing the second issue we have developed
a retrieval module
that ranks and retrieves Coherent Retrieval Elements (CREs) from the answer list of a native XML database.
We have shown that our CRE module is capable of retrieving answer elements with appropriate levels of retrieval granularity,
which means it could equally be applied with the VCAS retrieval task as it applies with the SCAS retrieval task.
Moreover, the CRE retrieval module can easily be used by other native XML databases,
since most of them output their answer lists in article order.

We have shown through the final results of our experiments that our hybrid XML retrieval system with the CRE retrieval module
improves the effectiveness of both retrieval systems and yields an effective content-and-structure XML retrieval.
However, this improvement is not as apparent as it is for content-only (CO) retrieval topics where no indication for
the granularity of the answer elements is provided~\cite{KluwerINEX03}.
The latter reflects the previous observation that the XML retrieval task should focus more on providing answer elements \emph{relevant} to
an information need instead of focusing on retrieving the elements that only satisfy the logical query constraints.

We plan to undertake the following extensions of this work in the future.

\begin{itemize}

\item Our CRE module
is currently not capable of comparing the ranking values of
CREs coming out of answer lists that belong to different articles.
We therefore aim at investigating whether or not additionally using Zettair as a means to rank the CREs coming out of different answer lists
would be an effective solution.

\item For further improvement of the effectiveness of our hybrid XML retrieval system, we also aim at investigating the optimal
combination of Coherent Retrieval and matching elements in the final answer list, which could equally be applied to CAS as well as to CO retrieval topics.

\end{itemize}

\section*{Acknowledgements}
We would like to thank Wolfgang Meier for providing assistance and useful comments with using eXist,
to Nick Lester, Falk Scholer and other members of the Search Engine Group at RMIT
for their support with using Zettair and to the anonymous reviewers for their useful suggestions.

\bibliographystyle{abbrv}
\bibliography{sigproc}  

\begin{thebibliography}{10}

\bibitem{INEX02_overview}
N.~Govert and G.~Kazai.
\newblock {Overview of the Initiative for the Evaluation of XML retrieval
  (INEX) 2002}.
\newblock In {\em Proceedings of the 1st Workshop of the {INitiative for the
  Evaluation of XML Retrieval (INEX)}, Dagstuhl, Germany, December 8-11, 2002},
  pages 1--17. ERCIM, 2003.

\bibitem{XRank}
C.~B. L.~Guo, F.~Shao and J.~Shanmugasundaram.
\newblock {XRANK: Ranked Keyword Search over XML documents}.
\newblock In {\em {Proceedings of the 2003 ACM SIGMOD international conference
  on on Management of data, San Diego, California, USA, June 9-12, 2003}},
  pages 16--27. ACM Press, 2003.

\bibitem{eXist}
W.~Meier.
\newblock {eXist: An Open Source Native XML Database}.
\newblock In {\em Web, Web-Services, and Database Systems. NODe 2002 Web- and
  Database-Related Workshops, Erfurt, Germany, October 7-10, 2002}, pages
  169--183. Lecture Notes in Computer Science 2593 Springer, 2003.

\bibitem{INEX03_overview}
S.~M. N.~Fuhr and M.~Lalmas.
\newblock {Overview of the INitiative for the Evaluation of XML retrieval
  (INEX) 2003}.
\newblock In {\em Proceedings of the 2nd Workshop of the {INitiative for the
  Evaluation of XML Retrieval (INEX)}, Dagstuhl, Germany, December 15-17,
  2003}, pages 1--11, 2004.

\bibitem{LuXistINEX03}
J.~Pehcevski, J.~Thom, and A.-M. Vercoustre.
\newblock {RMIT INEX Experiments: XML Retrieval using Lucy/eXist}.
\newblock In {\em Proceedings of the 2nd Workshop of the {INitiative for the
  Evaluation of XML Retrieval (INEX)}, Dagstuhl, Germany, December 15-17,
  2003}, pages 134--141, 2004.

\bibitem{KluwerINEX03}
J.~Pehcevski, J.~Thom, and A.-M. Vercoustre.
\newblock {Hybrid XML Retrieval: Combining Information Retrieval and a Native
  XML Database}.
\newblock Submitted for publication.

\bibitem{XSearch}
Y.~K. S.~Cohen, J.~Mamou and Y.~Sagiv.
\newblock {XSEarch: A Semantic Search Engine for XML}.
\newblock In {\em {Proceedings of 29th International Conference on Very Large
  Data Bases (VLDB), Berlin, Germany, September 9-12, 2003}}, pages 45--56.
  Morgan Kaufmann Publishers, 2003.

\bibitem{Pivot}
A.~Singhal, C.~Buckley, and M.~Mitra.
\newblock Pivoted document length normalization.
\newblock In {\em Proceedings of the 19th annual international ACM SIGIR
  conference on Research and development in information retrieval, Zurich,
  Switzerland, August 18-22, 1996}, pages 21--29. ACM Press, 1996.

\bibitem{INEX02_AnneMarie}
A.-M. Vercoustre, J.~A. Thom, A.~Krumpholz, I.~Mathieson, P.~Wilkins, M.~Wu,
  N.~Craswell, and D.~Hawking.
\newblock {CSIRO INEX experiments: XML Search using PADRE}.
\newblock In {\em Proceedings of the 1st Workshop of the {INitiative for the
  Evaluation of XML Retrieval (INEX)}, Dagstuhl, Germany, December 8-11, 2002},
  pages 65--72. ERCIM, 2003.

\bibitem{RossW}
R.~Wilkinson.
\newblock {Effective Retrieval of Structured Documents}.
\newblock In {\em Proceedings of the 17th Annual International Conference on
  Research and Development in Information Retrieval, Dublin, Ireland, July 3-6,
  1994}, pages 311--317. ACM Press/Springer, 1994.

\bibitem{MG99}
I.~Witten, A.~Moffat, and T.C.Bell.
\newblock {\em {Managing Gigabytes: Compressing and Indexing Documents and
  Images, 2nd edition}}.
\newblock Morgan Kaufmann Publishers, 1999.

\bibitem{wolff99-xpres}
J.~E. Wolff, H.~Fl\"orke, and A.~B. Cremers.
\newblock Searching and browsing collections of structural information.
\newblock In {\em Proceedings of IEEE Advances in Digital Libraries (ADL),
  Washington, DC, USA, May 22-24, 2000}, pages 141--150. IEEE Computer Society,
  2000.

\end{thebibliography}

\end{document}